\documentstyle[12pt]{article}

\begin{document}

\topmargin 0pt \oddsidemargin=-0.4truecm \evensidemargin=-0.4truecm 
\baselineskip=24pt 
\setcounter{page}{1} 
\begin{titlepage}     
\vspace*{0.4cm}
\begin{center}
{\LARGE\bf 
Search for Non-Standard Model  CP/T Violation at Tau-Charm Factory} 
\vspace{0.8cm}

{\large\bf  Tao Huang,~~Wei Lu~~and~ Zhijian Tao}

\vspace*{0.8cm}

{\em Theory Division, Institute of High Energy Physics, Academia Sinica\\
\vspace*{-0.1cm}
P.O.Box 918, Beijing 100039, China\\}
\end{center}
\vspace{.2truecm}     

\begin{abstract}
We systematically investigate the possibility of finding CP/T violation in the $\tau$ sector at 
Tau-Charm Factory. CP/T violation may occur at $\tau$ pair production process,  expressed as 
electric dipole moment,  and at $tau$ decay processes. By assuming that electric dipole moment as 
large as $10^{-19}$e-cm and CP/T violation effect orignating from $\tau$ decay as large as 
$10^{-3}$ are observable at Tau-Charm Factory, we studied all the possible extensions 
of the SM which are relevent for generating CP/T violation in $\tau$ sector.  And we 
pointed there are a few kind of  models, which are hopeful for generating such CP/T violation. 
For these models we 
consider all the theoretical and current experimental constraints and find that 
there exists some parameter space which will result in a measurable  CP/T violation.   
 Therefore we conclude that Tau-Charm Factory is a hopeful place to discover CP/T 
violation in $\tau$ sector.

\end{abstract}
\vspace{2cm}
\centerline{} 
\vspace{.3cm}
\end{titlepage}
\renewcommand{\thefootnote}{\arabic{footnote}} \setcounter{footnote}{0} 
\newpage

\section{Introduction}

The origin of CP violation has remained an unsolved problem since the
discovery of CP violation in K meson system a quarter ago\cite{Christ}.
Although the observed CP violation in K meson system can be accommodated in
the standard model (SM) of electroweak interactions by virtue of a physical
complex phase in the three by three Cabibbo-Kobayashi-Maskawa matrix (CKM) 
\cite{Kob}, it is not clear if CKM mechanism is really correct or the only
source for CP/T violation \cite{Lee}. To verify CKM mechanism one needs not
only the information on K meson mixing and decay but also that from the B
meson system or other systems. The main physical purpose of B factory is to
test the CKM mechanism. However even if CKM is the correct mechanism to
describe the CP violation in K and B meson mixing and decay, it is not
necessary that the CKM matrix is the only source of CP/T violation in the
nature \cite{Jar}. As pointed out by Weinberg \cite{Wei}, unless the Higgs
sector is extremely simple, it would be unnatural for Higgs-boson exchange
not to contribute to CP/T non-conservation. CKM matrix may explain the
observed CP violation in K meson system and possibly the CP violation in B
meson system, while other new sources of CP/T violation may occur everywhere
it can. In fact there are some physical motivations for people to seek the
new sources of CP/T violation. One motivation is from strong CP problem in
the SM \cite{Pec1}. For most of the scenarios to solve this problem they
need more complex vacuum structure and therefore new CP non-conservation
origin. Another motivation is from cosmology, most astrophysical
investigation shows that the additional sources of CP violation are needed
to account for the baryon asymmetry of universe at present \cite{Sha}. The
third motivation is from supersymmetry. Even in the minimal supersymmetrical
standard model (MSSM), there are some additional CP non-conservation sources
beyond the CKM matrix \cite{Hab}. Now the question is at what places the
possible new CP/T violation effects may show up and what is the potential to
search for those effects. In this work we are going to study systematically
on the possibility to find new CP/T violation effects at Tau-Charm factory
(TCF).

TCF is a very good place to test the SM and search new physics phenomena
because of its high luminosity and precision \cite{TCF}. Especially the $%
\tau $ sector is a good place to seek for non-SM CP/T violation effects
because in the SM CP violation in lepton sector occurs only at multi loop
level and is way below any measurable level in high energy experiments, only
non-SM sources of CP/T non-conservation may contribute and another reason is
that $\tau $ has abundant decay channels with sizable branching ratio, which
can be used to measure CP/T violation. Furthermore, the production-decay
sequences of $\tau $ pair by electron-positron annihilation is also favored.
The reason is as the following: (i) $\tau $ pair production by
electron-positron annihilation is a purely electroweak process and can be
perturbatively calculated; (ii) For the unpolarized electron-positron
collision, its initial state is CP invariant in the c.m. frame; (iii) when
the electron and/or positron beams are longitudinally polarized, the initial
state is still effectively CP even, which presents extra chances to detect
possible CP violation. To detect the possible CP/T violation, one can either
compare certain decay properties of $\tau ^{-}$ with corresponding CP/T
conjugations, or measure some CP/T-odd correlation of momentum or spin of
the final state particles from $\tau $ pair decay. These CP/T violating
observables can and should be constructed model independently, since
normally in non-SM these observables are not well predicted due to the
complexity and many free parameters. The sensitivity of the experimental
measurement on the possible CP/T violation is determined by the sensitivity
of the measurement on momentum, spin or other physical quantities of the
final state particles, from them the physical CP/T violating observables are
constructed. The better one can measure these quantities, the momenta, for
example, the smaller the CP violation phase can be reached. In TCF, it is to
expect about $10^7$ $\tau $ pair in one year, and the precision of
measurement on kinematic parameters at $10^{-3}$. The statistical and
systematic error can be around or below this level. Therefore generally a
CP/T violation phase as small as order of $10^{-3}$ can be reached at TCF 
\cite{TCF}. In a non-SM the CP/T violating phase may appear in various
stages of the process of production-decay chain, $e^{+}e^{-}\to \tau
^{+}\tau ^{-}\to final~~particles$. We sort them in three cases; (i) CP/T
violation is generated in the tree level production process, $e^{+}e^{-}\to
\gamma ,Z,X\to \tau ^{+}\tau ^{-}$, where X is some new Higgs or gauge
bosons, CP/T violating phase appears either in the propagator of X or in the
coupling to lepton pairs, and the simplest possibility is X being a neutral
Higgs in two or multi-Higgs doublets model. In this case the size of CP/T
violation is proportional to the interference between the X exchange and $%
\gamma $, Z exchange processes. Unfortunately for X being Higgs doublet this
interference term is proportional to the initial and final states fermion
masses $m_em_\tau $ as a result of chirality conservation. This factor along
contributes a suppression factor $m_e/m_\tau \sim 3\times 10^{-4}$ to all
CP/T violating observables in this kind of processes at TCF besides other
possible suppression factor, like the large mass of X, small coupling
between X and leptons. We conclude that it is hopeless to search for CP
non-conservation from the tree-level production process at TCF. (ii) CP/T
violation is also generated at production stage, but through loop level. The
most hopeful cases are that there may exist large electric or weak dipole
moment (EDM or WDM) for $\tau $ lepton, i.e. there are sizable CP/T
violation phase at the vertex $\tau ^{-}-\gamma ,Z-\tau ^{+}$. For this
situation the new physical particles beyond SM only appear as virtual
particles through loops and the size of CP violation is proportional to EDM
or WDM and is not suppressed by other factors, so the point is just whether
EDM or WDM of $\tau $ is large enough to be observed. Generally the
Lagrangian describing the CP/T violation in $\tau $ pair production related
to EDM and WDM is 
\begin{eqnarray}
L_{CP}=-1/2i{\bar \tau }\sigma ^{\mu \nu }\gamma _5\tau [d_\tau
^E(q^2)F_{\mu \nu }+d_\tau ^W(q^2)Z_{\mu \nu }] 
\end{eqnarray}
$F_{\mu \nu }$ and $Z_{\mu \nu }$ are the electromagnetic and weak field
tensors. The momentum transfer at TCF is around 4 GeV, and in LEP
experiments it is around the mass of Z boson. Therefore at TCF we expect the
contribution from WDM is a factor of $\frac{4m_\tau ^2}{M_Z^2}\simeq 2\times
10^{-3}$ smaller than the contribution from EDM, if EDM and WDM at the same
order of the magnitude. On the other hand the EDM term is less important at
LEP energy. That is the reason why the LEP data constrain more strictly on
WDM than EDM of $\tau $ \cite{ALE1,ALE}. We will neglect the WDM
contribution from now on in this work. (iii) It is possible that the CP/T
violation phase is small in the production process but it is relatively
large in the $\tau $ pair decay processes. The processes like $\tau $ to
neutrino plus light leptons or hadrons through some new bosons exchange at
tree level can contribute significantly to CP/T violation observables.
Obviously in this situation any CP/T violation effect from loop level is
negligible, since any loop effect is at least suppressed by a factor $\frac
1{16\pi ^2}\frac{m_\tau ^2}{M^2}$, where M is the mass of some new heavy
particles appearing in loops. This factor is smaller than $10^{-4}$ if M is
heavier than about 20 GeV.

Now let us recall that how one detects CP violation in $K$ meson decays: One
measures the partial widths for a decay channel and compares it with that
for the corresponding CP-conjugate decay process. Underlying such a
philosophy is the interference between a CP violating phase and a CP
conserving strong interaction phase, i.e. CP violation effect is only
manifested in the process with strong final state interaction. To observe
possible non-CKM CP violation effects in tau decays, however, one has to
invoke new methodology in the most cases. The basic reason is that both in
production vertex of $\tau $ pair (EDM of $\tau $) and in some tau decay
channels (like pure leptonic decay, $\pi \nu $, $\rho \nu $ decay channels
etc. ), there is no strong interaction phase, caused by hadronic final state
interaction, to interfere with possible CP violating phase. So far some
efforts have been made to investigate the CP/T violation effects in TCF.
Mainly those work are trying to find various ways to measure possible CP/T
violation. The simple and very useful method is to construct observables
which are CP/T-odd operators being made from momenta of final state
particles coming from $\tau $ pair decay or polarization vector of the
initial electron (or both electron and positron) beam \cite{Ber}. These
operators can be used very conveniently to test any CP/T violation from
either EDM of $\tau $ lepton or from the decay of the $\tau $ pair without
much model dependence. Some of the operators are constructed by considering
the reactions 
\begin{eqnarray}
e^{+}(p)+e^{-}(-p)\to \tau ^{+}+\tau ^{-}\to A(q_{-})+\bar B(q_{+})+X 
\end{eqnarray}
in the laboratory system, where $A(\bar B)$ can be identified as a charged
particle coming from the $\tau ^{-}(\tau ^{+})$ decay. Some CP/T-odd
operators (so CPT even, we will not consider CPT-odd operator in this work
since it is certainly much smaller violation effect) can be expressed as
following \cite{Ber}

\begin{eqnarray}
&O_1= 
{\hat p}\cdot \frac{{\hat q}_{+}\times {\hat q}_{-}}{|{\hat q}_{+}\times {\
\hat q}_{-}|}, \nonumber  \\ 
&T^{ij}=({\hat q}_{+}-{\hat q}_{-})^i\cdot \frac{({\hat q}%
_{+}\times {\hat q}_{-})^j}{|{\hat q}_{+}\times {\hat q}_{-}|}%
+(i\leftrightarrow j), 
\end{eqnarray}
where $\hat p,\hat q$ denote the unit momenta. If the initial electron
and/or positron beams are polarized, one can construct some more observables
making use of the initial polarization vector. For example a T violating
operator 
\begin{eqnarray}
O_2=\vec \sigma \cdot \frac{{\hat q}_{+}\times {\hat q}_{-}}{|{\hat q}%
_{+}\times {\hat q}_{-}|} 
\end{eqnarray}
can be constructed from the electron polarization vector $\vec \sigma $ and
momenta of final state particles. If there exists any sizable CP/T violation
from EDM of $\tau $ or in $\tau $ pair decay vertex, in principle the
experimental expectation values of these operators are nonzero. For EDM of $%
\tau $ lepton, $d_\tau $, the theoretical expectation values of these
operators are worked out and expressed only as a function of $d_\tau $ \cite
{Ana}. Since at TCF the precision of measurement for these operators are at $%
10^{-3}$ level, one expects to probe $d_\tau $ as small as $\displaystyle
\frac{10^{-3}}{2m_\tau }\simeq 10^{-17}$ e-cm. An example is the measurement
of $d_\tau $ or $d_\tau ^W$ in LEP experiment . Expectation value of $T^{ij}$
operator is directly related to $d_\tau $ \cite{ALE1}, 
\begin{eqnarray}
<T_{AB}^{ij}>=\frac{E_{cm}}ed_\tau C_{AB}diag(-1/6,-1/6,1/3). 
\end{eqnarray}
By the term diag means a diagonal matrix with diagonal elements given above, 
$E_{cm}$ is the energy at c.m. frame. The proportional constants $C_{AB}$
depend on the $\tau $ decay modes, but generally this constant is order of
one for all the decay models \cite{Ber}. The decay channels, which can be
measured in experiments, may be classified as $l-l$, $l-h$ and $h-h$
classes, here $l$ is the lighter leptons, $h$ is charged hadron like $\pi $, 
$\rho $ and $a_1$. Very impressively, if the initial electron (or both
electron and positron) is polarized, one may use the polarization
asymmetrized distribution. The distribution is defined as the differential
cross section difference between two different polarizations. With this
method, a $d_\tau $ as small as $10^{-19}$ e-cm can be reached at TCF \cite
{Ana}, this corresponds to a sensitivity of $10^{-5}$ of CP/T violation. Up
to now the best experimental bound on $d_\tau $ is from LEP experimental
data, which is used to exclude indirectly the $d_\tau $ as large as $%
10^{-17} $ e-cm \cite{ALE}, so two order of magnitudes improvement on $%
d_\tau $ measurement can be achieved at TCF.

Besides the CP/T-odd operator method, several other useful strategy were
proposed to test these violation in $\tau $ decay. 1) C. A. Nelson and
collaborators \cite{Nel} investigated systematically the feasibility of
using the so-called stage-two spin-correlation functions to detect possible
non-CKM CP violation in the tau-pair production-decay sequence and the
corresponding CP-conjugate sequence. The two-variable energy-correlation
distribution $I(E_A,E_B,\Psi )$, where $\Psi $ is the opening angle between
the final $A$ and $B$ particles, is essentially a kinematic consequence of
the tau-pair spin correlation which depends on the dynamics of $Z^0$ or $%
\gamma ^{*}\to \tau ^{-}\tau ^{+}$ amplitude, and of the $\tau ^{-}\to
A^{-}X_A$ and $\tau ^{+}\to B^{+}X_B$ amplitudes. By including $\theta _e$
and $\phi _e$ which specify the initial electron beam direction relative to
the final-state $A$ and $B$ momentum directions in the c.m. frame of $%
e^{-}e^{+}$ system, one obtains the so-called beam-referenced stage-two
spin-correlation function $I(\theta _e,\phi _e,E_A,E_B,\Psi )$. For the $%
\gamma ^{*}\to \tau ^{-}\tau ^{+}$ vertex, there are four complex helicity
amplitudes. Hence, the beam-referenced stage-two spin-correlation function
constructs four distinct tests for possible CP violation in $e^{-}e^{+}\to
\tau ^{-}\tau ^{+}$. To illustrate the discovery limit in using the
beam-referenced stage-two spin-correlation function, Goozovat and Nelson 
\cite{Nel2} calculated the ideal statistical errors corresponding to the
four tests. An advantage of detecting CP violation by use of the stage-two
spin-correlation function is that the model independence and amplitude
significance of the results is manifest. It is complementary to the greater
dynamical information that can be obtained through other approaches, such as
from higher-order diagrammatic calculations in the multi-Higgs extensions of
the SM. 2) Another strategy to test CP violation in the two-pion channels of
tau decay is due to Y.S. Tsai \cite{Tsa}, the basic ingredient of which is
to invoke a highly polarized tau-pair. Consider the tau-pair production by
electron-positron annihilation near threshold. If the initial electron and
positron beams are polarized longitudinally (along the same direction), the
tau-pair will be produced mainly in the $S$-wave, resulting in polarizations
of $\tau ^{\pm }$ both pointing in the same direction as that of the initial
beams. Such a polarization is independent of the production angle and the
corresponding polarization vector supplies us with an important block to
form products with the final particle momenta. By comparing such
polarization-vector-momentum products for a specific tau decay channel with
those for the corresponding CP-conjugate process, one can perform a series
of tests for possible CP violation effects in the tau decay. However, it is
impossible to detect a CP violation in the $\tau \to \pi \nu _\tau $ decay
without violating CPT symmetry. As for the two-pion channel, the existence
of a complex phase due to the hadronic final-state interactions, given by
the Breit-Wigner formula for the $P$-wave resonance $\rho $, enables
detecting possible non-CKM violation by measuring asymmetry of $({\bf w}%
\times {\bf q}_1)\cdot {\bf q}_2$ without violating the CPT symmetry, where (%
${\bf w}$ is the tau polarization vector and ${\bf q}_i$ (i=1,2) are the
final pion momenta). By limiting the weak interaction to be transmitted only
by exchange of spin-one and spin-0 particles, one can know that only the $S$%
-wave part of the amplitude for the exchange of the extra spin-1 particle
make contributions to CP violating observables. A very generic conclusion is
that unless two diagrams have different strong interactions phases, one
cannot observe the existence of weak phase using terms involving ${\bf %
w\cdot q_1}$. Tsai \cite{Tsai2} also points out that T violation can not be
detected in the pure leptonic decay without detecting the polarization of
the decay lepton. Because it is impossible to construct T-odd operator by
the momenta of the initial and final state particles in pure leptonic three
body decays. This also implies that with CPT symmetry, one can not detect CP
violation in $\tau $ decay processes with unpolarized $\tau $. On the other
hand, however, with polarized initial electron and positron beams, one can
construct T-odd operators using the momenta and polarization vector of $\tau 
$ and the decay lepton. Therefore polarization of initial electron and
positron is very desirable for detecting of CP/T violation at TCF. 3) As for
the $\tau \to (3\pi )\nu _\tau $ decay, it can proceed either via $J^P=1^{+}$
resonance $a_1$ and the $J^P=0^{-}$ resonance $\pi ^{\prime }$. Choi,
Hagiwara and Tanabashi \cite{Cho} investigated the possibility that the
large width-mass ratios of these resonances enhance CP-violation effects in
the multi-Higgs extensions of the SM. To detect possible CP-violation
effects, these authors compare the differential decay width for the $\tau
^{-}\to \pi ^{+}\pi ^{-}\pi ^{-}\nu _\tau $ with that for the corresponding
CP-conjugate decay process. To optimize the experimental limit, they
suggested considering several CP-violating forward-backward asymmetry of
differential decay widths, with appropriate real weight functions. 4) To
probe possible CP-violating effects in the tau decay with $K^{-}\pi ^{-}\pi
^{+}$ or $K^{-}\pi ^{-}K^{+}$ final states, Kilian, K\"orner, Schilcher and
Wu \cite{Wu} partitioned the final-state phase space into several sectors
and constructed some asymmetries of the differential decay widths. As a
result, they showed that T-odd triple momentum correlations are connected to
certain asymmetries, so their non-vanishing would indicate a possible
non-CKM CP violation in the exclusive semileptonic $\tau \to $three
pseudoscalar-meson decays.

With these knowledge and results obtained in the previous papers in mind,
now the crucial question, which is also the motivation of this work, is
whether for CP/T violation appearing in EDM close to $d_\tau \sim 10^{-19}$
e-cm and CP/T violation effects in $\tau $ decay as $10^{-3}$ are possible
values theoretically. If for all possible extensions of the SM, which people
can visualize now, with natural parameter choice, these values are much
smaller than the theoretically predicted ones, then the effort to search for
such small CP/T violation signal at TCF would be not much meaningful, at
least from the theoretical point of view. In this paper we are trying to
answer this question by investigating various possible mechanisms for
generating large EDM of $\tau $, CP/T violation in $\tau $ decay. This paper
is organized as the following. In section 2 we review the generation of EDM
of $\tau $ lepton in various popular beyond standard models and stress on
what models can produce possible large EDM of $\tau $. Following the
discussion of EDM, in the section 3 we concentrate on CP/T violation effects
from $\tau $ decay in the beyond standard models. The last section is
reserved for some further discussion, and the conclusion on the possibility
of finding CP/T violation at TCF is given.

\section{EDM of $\tau$ lepton}

EDM of the lepton $d_l$ is a dimension-5 operator. It can only be generated
from the loop level. Because this operator changes the chirality of the
lepton, it must be proportional to a fermion mass. In the SM EDM of lepton
is generated from three loop diagrams and is proportional to lepton mass
itself, so $d_l$ is very small \cite{Khr}. However generally $d_l$ can be
produced from one loop diagrams in beyond standard model. At one loop level,
the $d_l$ can be expressed as 
\begin{eqnarray}
d_l\sim \frac{e\lambda }{16\pi ^2}\frac{ M_F}{V^2}\sin\phi\sim 10^{-18}( 
\frac{\lambda}{1} )(\frac {M_F}{100GeV})(\frac{100GeV}V)^2\sin\phi ~~\rm{e-cm} 
\end{eqnarray}
where $M_F$ is some fermion mass, $V$ is a large scale from intermediate
states in the loops and $\lambda $ denotes other couplings. $\phi$ is a CP/T
violation phase. In the following part we assume maximal CP/T violation
phase, i.e. $\sin\phi\simeq 1$. From this equation one sees that $d_l$ can
be at most as large as $10^{-18}-10^{-19}$ e-cm if $\lambda$ is between $%
1.0-0.1 $. Since $V$ is a scale around or larger than weak scale, in order
to obtain large $d_l$, $M_F$ must be a large fermion mass such as t quark
mass or new heavy fermion masses. For example if M is the $\tau $ mass then $%
d_\tau $ is smaller than $10^{-20}$ e-cm which is not detectable at TCF.
Same is true for the scale $V$ . If $V$ is at TeV scale $d_l$ is smaller
than $10^{-20}$ e-cm. Although, in principle, $d_\tau $ is possibly as large
as $10^{-19}$e-cm , one has to avoid too large EDM of electron $d_e$ at the
same time. Current experimental upper limit on $d_e$ is about $10^{-26}$%
e-cm. This is a very strong constraint especially when one is expecting
large $d_\tau $. So in any beyond standard model, two requirements must be
satisfied in order to obtain measurable $d_\tau $. The first one is that the
model must provide $d_\tau $ at one loop level and $d_\tau $ is not
suppressed by a small fermion mass term, the fermion mass term should be a
top quark mass, supersymmetric partner of bosons or other exotic fermion
masses. The second one is that the predicted $d_e$ associated with large $%
d_\tau $ is below its current experimental bound. These two conditions
altogether exclude most of beyond standard models which can provide large
enough $d_\tau $ observable for TCF. We will see from the following
discussion that many beyond standard models do not satisfy the two
requirements.

Usually EDM of lepton is generated from one loop diagrams in extension of
the SM. Fig. 1 is a typical one loop diagram for the lepton EDM. The virtual
particles are scalar or vector boson $S$ and fermion $F$ in the loop. Photon
is attached to the charged intermediate particles. The $d_l$ from this
diagram is approximately proportional to the fermion mass $M_F$ and it is
divided by a scale $V$, which is larger or equal to $M_F$. Besides, there
are two more couplings at the vertex $l-S-F$. In a practical model there
could be many possible virtual bosons and fermions in the loop, but we only
consider the dominant contribution here as an order of magnitude estimation.
The diagram in Fig. 1 is evaluated as 
\begin{eqnarray}
d_i/Q\simeq \frac{|\lambda _i\lambda ^{\prime *}_i|}{16\pi ^2}\frac{M_F}{V^2}
\xi\sin\phi 
\end{eqnarray}
where $i=e,~\mu ,~\tau $ denotes three generation leptons and $Q$ is the
electric charge of the virtual particles. $\xi$ is an order of one factor
from the loop integral. Eq. (7) is true up to a factor of order one. And
there should be a logarithmic dependence on $\frac{M_F}V$ in $\xi$, which is
slowly varying.

In order to obtain measurable $d_\tau $ and avoid too large $d_e$, one needs
a large $M_F$ as discussed before and $\lambda $, $\lambda ^{\prime }$ must
be around order of one for $\tau $ but much small (smaller than about $%
10^{-3}$) for electron. We systematically investigate and review most of the
popular extensions of the standard model and point out that the following
type of models can fulfill the requirements.

{\bf {Scalar leptoquark models}}\cite{Dav} \hskip 0.5 cm CP violation effect
in $\tau$ sector for the models are recently discussed extensively by some
authors \cite{Cho,Bar}. It is particularly interesting for generating a
large $d_l$. These are the models which do not need to introduce additional
fermion. Because the top quark mass is large, it is possible to generate a
large $d_{\tau}$ through coupling of $\tau$, top quark and the corresponding
leptoquark. $d_e$ could be small enough due to the coupling of electron, top
quark and leptoquark is independent of that for $d_{\tau}$. So long as there
is a relative large hierarchy for the couplings for different generations,
the two requirements can be satisfied.

There are five types of scalar leptoquarks which can couple to leptons and
quarks. We denote them by $S_1$, $S_2$, $S_3$, $S_4$ and $\vec{S_5}$. Their
quantum numbers under standard gauge group transformation are $(3, 2, \frac{%
7 }{3})$, $(3, 1, -\frac{2}{3})$, $(3,2,\frac{1}{3})$, $(3,1,-\frac{7}{3})$
and $(3, 3, -\frac{2}{3})$ respectively. The Yukawa coupling terms are
therefore given by 
\begin{eqnarray*}
&L_1=(\lambda_1^{ij} 
{\bar Q}_{Li}i\tau_2E_{Rj}+\lambda^{\prime ij}_1{\bar U}_{Ri}l_{Lj})S_1+h.c. \nonumber
\\ &L_2=(\lambda_2^{ij} 
{\bar Q}_{Li}i\tau_2l^c_{Lj}+\lambda^{\prime ij}_2{\bar U}%
_{Ri}E^c_{Rj})S_2+h.c. \nonumber \\& L_3=\lambda_3^{ij} 
{\bar D}_{Ri}l_{Lj}S_3+h.c.  \nonumber \\ &L_4=\lambda_4^{ij} 
{\bar D}_{Ri}E^C_{Rj}S_4+h.c. \nonumber \\ &L_5=(\lambda_5^{ij} {\bar Q}_{Li}i\tau_2 
\vec{\tau} l^c_{Lj})\cdot\vec{ S_5}+h.c. 
\end{eqnarray*}
Here $l_L$ and $Q_L$ are lepton and quark doublets respectively, $U_R$, $D_R$
and $E_R$ are singlet quark and lepton respectively. Individually only $S_1$
and $S_2$ contribute to the EDM of lepton.

$\xi $ factor in Eq. (7) is evaluated as $\xi =\frac 23ln\frac{M_F^2}{V^2}+ 
\frac{11}6$ \cite{Bar}. Currently the constraints on mass and coupling of
leptoquark are relatively weak \cite{DAT}. For leptoquark coupled only to
third generation, its lower mass bound is about 45 GeV with order of unit
coupling \cite{DAT}. This bound is from a leptoquark pair production from
LEP experiments. On the other hand with the leptoquark mass at weak scale,
the coupling is very weakly bounded too. In fact the coupling could be as
large as order of one. If we take $\lambda ^{33}$ , $\lambda ^{\prime
}{}^{33}$ as 0.5 and the mass of leptoquark as 200GeV and assume maximal
CP/T violation phase, we estimate that $d_\tau \simeq 2\times 10^{-19}$
e-cm, while $d_e$ is determined by other coupling components, so a small $%
d_e $ is not necessary in conflict with a large $d_\tau $ in this model.

{\bf Models with the fourth generation or other exotic lepton}\hskip 0.5cm
The SM with fourth generation is another possible model to generate a large $%
d_{\tau}$. The heavy fourth generation leptons may play a role of the heavy
fermion F in the loop. However it is well known that if the fourth
generation exists, it must satisfy the constraints from LEP experiments \cite
{LEP}. Here we propose a realistic model for this purpose.

Besides the fourth generation fermions, we also introduce a right-handed
neutrino $\nu_R$ and a singlet scalar $\eta^-$ with one unit electric charge 
\cite{Zee}. The new interaction terms are 
\begin{eqnarray}
L=\lambda_{ij}l_i^Ti\tau_2l_j\eta^-+\lambda^{\prime}_{i}E_{Ri}^T\nu_{R}%
\eta^-+ M_R\nu_R^T\nu_R+M_i^D{\bar\nu}_{Li}\nu_R+h.c. 
\end{eqnarray}
where $\lambda_{ij}$ is antisymmetric due to the Fermi statistics. $M^D$ is
the Dirac neutrino mass from standard Higgs vacuum expectation value. In
this model three light neutrinos remain massless and the fourth neutrino is
massive \cite{Li}. The constraints from LEP experiments and other low energy
data can be satisfied so long as $M_R$ is at weak scale or up and $M_i^D$ is
not much smaller than $M_R$. In the one loop diagram contribution to $d_\tau 
$, $\eta ^{-} $ appears as the scalar S. The fermion line is two massive
neutrinos $\nu _4$ and $\nu _H$ in the mass basis and they are related to
each other, 
\begin{eqnarray}
\nu _{L4}=\cos \theta \nu _4-\sin \theta \nu _H  \nonumber\\ 
\nu _R=\sin \theta \nu _4+\cos \theta \nu _H 
\end{eqnarray}
We assume $\nu _4$ is the lighter neutrino and the dominant contribution is
from either $\nu_H$ or $\nu _4$ depending on whether $\nu _H$ is heavier
than the mass of $\eta $, $M_\eta $ . $d_\tau $ is evaluated as in (7) with $%
M_F=M_H\cos \theta \sin \theta $ and $V\simeq M_\eta $ if $M_\eta \geq M_H$;
with $M_F=M_{\nu _4}\cos \theta \sin \theta $ and $V\simeq M_H$ if $M_\eta
\leq M_H$. Choosing $\lambda _{34}=\lambda _3^{\prime }=1.0$ and $M_F=50$
GeV, $V=200$ GeV, we have the numerical result $d_\tau \simeq 10^{-19}$
e-cm. Also in this model a hierarchy on the coupling $\lambda $ and $\lambda
^{\prime }$ for different generation is needed to keep small enough $d_e$,
i.e. $\lambda _{34}>>\lambda _{14}$ and $\lambda _3^{\prime }>>\lambda
_1^{\prime }$.

Existence of exotic leptons provide another possibility to generate a
measurable $d_{\tau}$. It can be realized in horizontal models \cite{Bar1}.
With only three standard leptons, it is impossible to obtain large enough $%
d_{\tau}$, because the largest fermion mass in the loop is $m_{\tau}$.
However, with some new heavy leptons this model can provide a large $%
d_{\tau} $. The constraints from low energy data can be avoided if one
assumes that the horizontal interaction is strong between $\tau$ and the
exotic lepton, but it is much weaker in other sectors. Similar result on $%
d_{\tau}$ as for the case with the fourth generation can be obtained.

Finally, we should point out that for our purpose it is clear that some new
exotic heavy leptons are needed in the new physics models, however even
though there exists some kind of models with some new heavy leptons, they
are able to generate $d_l$ only from two loop diagrams \cite{Fab}, so they
may result in interesting $d_e$ but not $d_\tau $.

{\bf Generic MSSM } \hskip 0.5pc Generic MSSM contains 63 parameters not
including the parameters in the non-SUSY SM. Ferminic superpartners of the
ordinary bosons can be the heavy fermions in the loop diagrams for $d_l$. It
provides some new sources for CP/T violation. It is well known that the
electron and neutron can acquire large EDM \cite{Pol} in this model. In
fact, in order to obey the experimental bounds on $d_n$ and $d_e$, some
parameters in the model are strongly restricted \cite{Bab}. For $d_l$
generation, it is dominated by photino mediated one loop diagram. Both left-
and right-handed sleptons also appear in the loop. The contribution to $d_l$
from this diagram is proportional to left- and right-handed slepton mixing
matrices $M_{LR}=(A_l-\mu \tan \beta )M_l$. $A_l$ is the matrix of
soft-SUSY-breaking parameters that appears in the SUSY Yukawa terms of
slepton coupling to Higgs doublet. Here $M_l$ is diagonal mass matrix of
lepton mass. Usually it is assumed that $A_l$ is diagonal and the diagonal
elements are not much different for different generation, for example in
supergravity inspired model $A_l$ is universal for three generation \cite
{Hab}, therefore one can get $d_\tau /d_e\simeq m_\tau /m_e$. Using the
experimental limit $d_e\leq 10^{-26}$ e-cm, one concludes that $d_\tau \leq
4\times 10^{-23}$ e-cm \cite{Mah}. However in the generic MSSM all the
elements of $A_l$ are free parameters, so the above constraint is not
necessarily true. For example if for some unknown reason the 33 component of 
$A_l$ is much larger than other elements, and $\mu $ term is much smaller
than SUSY breaking scale, then $d_\tau $ still can be larger than $10^{-22}$
e-cm and $d_e$ is in the allowed region. In this case $d_\tau $ can also be
expressed as Eq. (7), but with $M_F={\tilde m}_\gamma $, $V={\tilde m}_\tau
^2/M_{LR}$, $\lambda _{33}=\lambda _{33}^{\prime }=e$ and $\phi =arg(M_{LR}^2%
{\tilde m}_\gamma )$. The loop integral $\xi $ was four times the function
calculated some years ago in dealing with $d_e$ in MSSM known as
Polchinski-Wise function \cite{Wis}. Here ${\tilde m}_\gamma $ and ${\tilde m%
}_\tau $ are photino and the third slepton masses respectively. We estimate
that $d_\tau \simeq 10^{-19}$ e-cm with ${\tilde m}_\gamma =100$ GeV and $%
V=200$ GeV.

As for other popular extensions of the SM, we would like to point out here,
though they have some new sources of CP/T violation, they can not offer a
observable $d_\tau $ at TCF. These include multi-Higgs doublet model (
including two Higgs doublet model) \cite{Lee,Wei2}, Left-Right symmetric
model \cite{Pat}, mirror fermion model \cite{Don} and universal soft
breaking SUSY model \cite{Hab} etc. In multi-Higgs doublet model electron 
\cite{Barr} and neutron \cite{Wei} may obtain a large EDM close to current
experimental bounds through two loop diagrams, but $d_\tau $ generated in
the model is quite below the TCF observable value. The reason is that $%
d_\tau $ is proportional $m_\tau $, but not a large fermion mass. We
estimate $d_\tau \leq 4\times 10^{-21}$e-cm \cite{Son} that in this model.
For Left-Right symmetric model, Nieves, Chang and Pal \cite{Nie} find that
the upper bound for $d_\tau $ is $2.4\times 10^{-22}$e-cm. It is the right-
or left-handed gauge boson in the loop as the role of $S$ particle, while
right-handed neutrino is the virtual fermion particle in the loop. $d_\tau $
in this model is proportional to left- and right-handed gauge boson mixing
angle. Though it is not suppressed by the small fermion mass ( $M_F$ is a
large right-handed neutrino mass), the mixing angle is constrained to be
smaller than $0.004$ \cite{Don2} from purely non-leptonic strange decays. It
leads to about three order of magnitude suppression. In the mirror fermion
model, standard gauge bosons couple to ordinary lepton and the mirror lepton
with a mixing angle. It is $Z$ and $W$ bosons in the one loop diagrams, the
heavy fermion line is the mirror lepton. However the mixing angle in this
model is constrained by various experiments \cite{Lan}, and most stringently
by LEP data on $Z\to \tau ^{+}\tau ^{-}$ \cite{Bha}. The constraint from LEP
data on the mixing angle is less than about 0.3. The resulting bound is $%
d_\tau \leq 2.1\times 10^{-20}$e-cm, which is a few times smaller than TCF
measurable value. As we have mentioned above in the universal soft breaking
SUSY model, $d_\tau \leq 4\times 10^{-23}$e-cm due to the constraint on $d_e$%
. The only alternative situation is discussed above on Generic MSSM in this
section.

\section{CP/T violated $\tau$ decays}

As we have pointed out in the introduction, CP/T violation effects in $\tau $
decays, if observed, must occur at tree level diagrams. That is the
interference between the SM $\tau $ decay processes and new tree level
processes of $\tau $ decays, in which CP/T violation phases appear at the
interaction vertexes, provides the information of CP/T violation in the $%
\tau $ sector. Feynman diagrams of these processes can be shown as in the
Fig. 2, where $f_i$, $f_j$ and $f_k$ are light
fermions. X is a new particle ( scalar or vector boson) which mediates CP/T
violating interaction. The size of CP/T violation is always proportional to
the interference of the tree level diagrams. We denote the amplitudes for
these diagrams as $A_1$ for W boson exchange diagram, $A_2$ for other X
boson exchange diagrams. The size of CP/T violation in the $\tau $ decay can
be characterized by a dimensionless quantity 
\begin{eqnarray}
\epsilon =\frac{Im(A_1^{*}A_2)}{|A_1|^2+|A_2|^2} 
\end{eqnarray}
Practically physical quantity expectation values which are used to reflect
CP/T violation, like the expectation values of CP/T-odd operators,
difference of a partial decay widths of a $\tau ^{-}$ decay channel and its
conjugate $\tau ^{+}$ decay channel, are model dependent and generally quite
complicate. It needs the detailed information of the new physics model and a
lot of parameters enter into the expression. This makes it a very much
involved work to write down these quantities in a specific model beyond the
SM. And the exact CP/T violation quantity expression written down from a
model should be different from the $\epsilon $ defined above. However as a
simple and reasonable estimation, the quantity $\epsilon $ in Eq. (11) can
be used as an indication of how large of CP/T violation may happen at
various $\tau $ decays. Moreover, the amplitude $A_2$ is usually much
smaller than $A_1$ because so far all the experimental data agree with the
SM prediction very well. So $A_2$ term in the denominator can be neglected.
Using $A_1$ as the amplitude from W boson exchange and $A_2$ from the new
boson X exchange, we estimate its size, 
\begin{eqnarray}
\epsilon\sim (4\sqrt{2}G_F)^{-1}\frac{Im(\lambda\lambda^{\prime}{}^*)}{M_X^2}
\end{eqnarray}
Here $G_F$ is Fermi constant and $\lambda$, $\lambda^{\prime}$ are couplings
in $A_2$. From Eq. (12) one sees that the size of CP/T violation is
determined by the parameter $\frac{Im(\lambda\lambda^{\prime}{}^*)}{M_X^2}$.
For different models, this parameter is constrained by some other physical
processes. So the possible size of CP/T violation depends on the parameter
region which is restricted in a specific model.

In Fig. 2 the final state fermions can be a pair of leptons and quarks
besides $\nu _\tau $. It corresponds to pure leptonic and hadronic decays
respectively. At the quark level, the diagrams with a pair of quarks in the
final states denote an inclusive process, it includes all possible hadronic
channels originated from quark pair hadronization. Some of the useful
hadronic final states like $2\pi $, $3\pi $, $K\pi $, $K\pi \pi $, $KK\pi $
and $\rho $, $a_1$ can be used to measure the properties of $\tau $.
However, it is often difficult to make a reliable quantitative prediction
for CP/T violation in exclusive hadronic decay modes, because of the
uncertainty in the hadronic matrix elements. On the other hand, for the
inclusive cases, one may make a more reliable quantitative estimation due to
the fact that one has no need to deal with the hadronization of quarks in
this case. In addition, QCD correction should not change the order of the
tree level diagram evaluation as the energy scale for $\tau $ decay
processes is around 1GeV. In this section we only deal with the diagrams
containing quark pair inclusively, So the CP/T violation size we estimate
below is for all the possible hadronic decay channels. In the last section
we will comment on our results in exclusive processes. Because of the scale
of $\tau $ mass, its decay products can only be neutrino, electron, muon and
hadrons containing only light u,d, s quarks as other heavy quarks are
kinematically forbidden. Therefore there are not many possibilities for X
particle being the candidate for mediating CP/T violation in the Fig. 2. In
fact all the possible choices are the following: X being leptoquark, charged
Higgs singlet, doublet and triplet, and double charged singlet. Now we come
to discuss these different cases separately.

{\bf Scalar leptoquark models} \hskip 1cm At tree level it is obvious that
only $S_1$, $S_2$ and $\vec S_5$ contribute to $\tau $ decays. There are two
type of decay processes at quark level, $\tau \to \nu _\tau {\bar u}d$ and $%
\tau \to \nu _\tau {\bar u}s$. The $\epsilon $ parameter is determined by $%
\lambda ^{31}{\lambda ^{\prime }}^{31}{}^*$ and $\lambda ^{32}{\lambda
^{\prime }}^{31}{}^{*}$ for these two type of decays respectively in model 1
and 2 in Eq. (8). For model 5 there is CP/T violation effect only in the
second type process, which is determined by $\lambda ^{32}{\lambda ^{\prime }%
}^{31}{}^{*}$. A direct constraint on these parameters can be obtained
through comparing the theoretical value $\Gamma ^{th}(\tau \to \pi \nu _\tau
)=(2.480\pm 0.025)\times 10^{-13}$ GeV and the measurement value of $\Gamma
^{exp}(\tau \to \pi \nu _\tau )=(2.605\pm 0.093)\times 10^{-13}$ GeV \cite
{Mar}. Assuming that real and imaginary part of the coupling $\lambda {%
\lambda ^{\prime }}{}^{*}$ are approximately equal, one has from $\tau\to
\pi \nu_{\tau}$ \cite{Cho} 
\begin{eqnarray}
\frac{|Im(\lambda ^{31}{\lambda ^{\prime }}^{31}{}^{*})|}{M_X^2}\sim \frac{%
|Re(\lambda ^{31}{\lambda ^{\prime }}^{31}{}^{*})|}{M_X^2}<3\times
10^{-6}GeV 
\end{eqnarray}
at $2\sigma $ level for model one and two. And from $\tau \to \ K\nu _\tau $
a similar result can be obtained for all the three models. Using the
theoretical value $\Gamma ^{th}(\tau \to K\nu_{\tau} )=(0.164\pm
0.036)\times 10^{-13}$ GeV \cite{Mar,Mar1} and the measurement value $\Gamma
^{exp}(\tau \to K\nu _\tau )=(0.149\pm 0.051)\times 10^{-13}$ GeV for the $%
\tau \to K\nu _\tau $ decay width we obtain 
\begin{eqnarray}
\frac{|Im(\lambda ^{32}{\lambda ^{\prime }}^{31}{}^{*})|}{M_X^2}\sim \frac{%
|Re(\lambda ^{32}{\lambda ^{\prime }}^{31}{}^{*})|}{M_X^2}<7\times
10^{-6}GeV 
\end{eqnarray}
at $2\sigma $ level. This constraint is less stringent due to the large
uncertainties in $\Gamma ^{exp}(\tau \to K\nu _\tau )$. With these
constraints, one estimates the upper bound of $\epsilon $ value for the two
type of processes as 
\begin{eqnarray}
\epsilon (\tau^- \to \nu _\tau {\bar u}d)\simeq (4\sqrt{2}G_F)^{-1}\frac{%
Im(\lambda ^{31}{\lambda ^{\prime }}^{31}{}^{*})}{M_X^2}\leq 4\times 10^{-2} 
\end{eqnarray}
and 
\begin{eqnarray}
\epsilon (\tau^- \to \nu _\tau {\bar u}s)\simeq (4\sqrt{2}G_F)^{-1}\sin
\theta _C\frac{Im(\lambda ^{32}{\lambda ^{\prime }}^{31}{}^{*})}{M_X^2}\leq
2\times 10^{-2} 
\end{eqnarray}
where $\theta _C$ is Cabibbo angle. $\epsilon (\tau^- \to \nu _\tau {\bar u}%
s)$ is proportional to $\sin \theta _C$ and is smaller than $\epsilon (\tau
\to \nu _\tau {\bar u}d)$ because this process is Cabibbo suppressed, even
though the coupling is less constrained than that of Cabibbo unsuppressed
process. From this estimation we expect CP/T violation in these models could
be large enough for TCF or in the other words TCF data can put stronger
direct restriction on the parameters of the model. However, if one assumes
that all the couplings $\lambda $ and $\lambda ^{\prime }$ are at the same
size irrespective of the generation indexes, then much more stringent bounds
exist. These bounds are obtained from experimental bounds of $Br(K_L\to\mu
e) $, $Br(\pi\to e\nu_e(\gamma))$, $Br(\pi\to \mu\nu_{\mu}(\gamma))$ and $%
\Gamma(\mu Ti\to e Ti)/ \Gamma(\mu Ti\to capture)$ \cite{Cho}. They are
generally about five order of magnitude smaller than the direct bounds.
Therefore the size of CP/T violation is $\epsilon \le 4\times 10^{-7}$ which
is far below the capability of TCF.

{\bf Multi-Higgs doublet models (MHD)} \hskip 1cm With the natural
suppression of flavor changing neutral current, it is necessary to have more
than two Higgs doublets, so that there are at least two physical charged
Higgs particles. CP/T violation may generally happen through the mixing of
these charged Higgs particles. We consider a multi-Higgs doublet model, say,
n Higgs doublets. In this model there are 2(n-1) charged and (2n-1) neutral
physical scalars. Since only the Yukawa interactions of the charged scalars
with fermions are relevant for our purpose. Following Grossman \cite{Gro} we
write down the Yukawa interactions in fermion mass eigenstates as 
\begin{eqnarray}
L_{MHD}=\sqrt{2\sqrt{2}G_F}\Sigma _{i=2}^n[X_i({\bar U}_LVM_DD_R)+Y_i({\bar U%
}_RM_UVD_L+Z_i({\bar l}_LM_EE_R)]H_i^{+}+h.c. 
\end{eqnarray}
Here $M_U$, $M_D$ and $M_E$ denote the diagonal mass matrices of up-type
quarks, down type quarks and charged leptons respectively. $V$ is KM matrix. 
$X$, $Y$ and $Z$ are complex couplings which arise from the mixing of the
charged scalars and CP/T violation in $\tau $ decay processes is due to
these couplings. How large is the $\epsilon $ for various $\tau $ decay
channels depends on the values of these parameters. More precisely, in the
pure leptonic decays the size of CP/T violation is determined by $%
Im(Z_iZ_j^{*})$ with $i\not =j$ and in hadronic decays it is determined by $%
Im(X_iZ_j^{*})$ and $Im(Y_iZ_j^{*})$. The three combinations of parameters
are constrained by various experiments \cite{Gro}. The strongest constraint
on $Z$ is from $e-\mu $ universality in $\tau $ decay, which gives $|Z|\le
1.93M_H$GeV$^{-1}$ for Higgs mass $M_H$ around 100GeV. $Im(XZ^{*})$ is
bounded from above from the measurement of the branching ratio $Br(B\to
X\tau \nu _\tau )$, $Im(XZ^{*})\le |XZ|\le 0.23M_H^2$GeV$^{-2}$ if $M_H\le
440$ GeV. Finally a upper bound is given as $Im(YZ^{*})\le |YZ|\le 110$ from
the experimental data of the process $K^{+}\to \pi ^{+}\nu {\bar \nu }$
This bound is obtained for t quark mass at 140 GeV \cite{Gro} and $M_H=45$%
GeV, however for a different $M_H$, say 100GeV, this bound is expected not
to change much. With these bounds we can estimate CP/T violation size of $%
\tau $ leptonic and hadronic decays. For the leptonic decay $\tau \to \mu
\nu {\bar \nu }$, we have the quantity 
\begin{eqnarray}
\epsilon \simeq \frac 12\frac{Im(ZZ^{*})m_\mu m_\tau }{M_H^2}\cdot \frac{%
m_\mu }{m_\tau }=\frac 12\frac{m_\mu ^2}{M_H^2}Im(ZZ^{*})\leq 2\times
10^{-2}. 
\end{eqnarray}
Here the additional factor $\frac{m_\mu }{m_\tau }$ comes from the
interference of left- and right-handed muon lines in the final states. So we
expect that CP/T violation effect in the process $\tau \to e\nu {\bar \nu }$
is suppressed by a factor $m_e/m_\mu $ and is negligible. For the hadronic
decay $\tau \to {\bar u}d\nu $ we have 
\begin{eqnarray}
\epsilon \simeq \frac 12\frac{m_d{\bar m}_d}{M_H^2}Im(XZ^{*})\leq 3\times
10^{-4}, 
\end{eqnarray}
With the current $d$ quark mass $m_d=7$ MeV and the dynamical $d$ quark mass 
${\bar m}_d=300$ MeV . For hadronic decay $\tau \to {\bar u}s\nu $ similar
result is obtained 
\begin{eqnarray}
\epsilon \simeq \frac 12\frac{m_s{\bar m}_s}{M_H^2}Im(XZ^{*})\leq 1.5\times
10^{-3} 
\end{eqnarray}
Here we use current and dynamical $s$ quark masses as 150 MeV and 400 MeV
respectively. In summary, in multi-Higgs doublet model CP/T violation effect
is possibly as large as order of $10^{-3}$ for exclusive hadronic decays and
It could be even close to $10^{-2}$ in pure leptonic decay to $\mu $ and
neutrinos.

{\bf Other extensions of the SM for pure leptonic decays}\hskip 1cm Besides
leptoquark and Higgs doublet, there are three other kind of scalars which
can couple to leptons. We denote $l$ as a lepton doublet and $E$ as a
singlet lepton. Two $l$ can combine to a charged singlet or a triplet. Two $%
E $ can combine to a double charged singlet. Corresponding to these three
cases one can introduce a charged singlet scalar $h^{-}$, triplet scalar $%
\Delta $ and double charged scalar $K^{--}$. However $K^{--}$ only induce a
lepton family- number-violating process $\tau \to 3l$. There is no diagram
corresponding SM contribution, so there is no CP/T violation mediated by
this particle. Also the branching ratio ($\leq 10^{-5}$) for this decay is
much smaller than TCF reachable CP/T violation precision $10^{-3}$. In
principle if there exists more than one $h$ or $\Delta $, CP/T violation can
be induced by the interference of the W exchange diagram and $h$ or $\Delta $
exchange diagram in the process $\tau \to l{\bar \nu }\nu $ with $l=e,\mu $.
Now let us discuss these two possibilities in details. We can write down the
new interaction terms which couple the new scalar particles to leptons as
the following 
\begin{eqnarray}
L_h=\frac 12f_{ij}l^T{}_iCi\tau _2l_jh+h.c. 
\end{eqnarray}
\begin{eqnarray}
L_\Delta =\frac 12g_{ij}l^T{}iCi\tau _2{\vec \tau }l_j{\vec \Delta }+h.c., 
\end{eqnarray}
where $C$ is the Dirac charge conjugation matrix and $f_{ij}$ is
antisymmetric, $g_{ij}$ is symmetric due to Fermi statistics. $\epsilon $
parameter for these singlet and triplet models are given by, 
\begin{eqnarray}
\epsilon _h\simeq (4\sqrt{2}G_F)^{-1}\frac{Im(f_{\tau l}f_{l\tau }^{*})}{%
M_h^2} 
\end{eqnarray}
in singlet model and 
\begin{eqnarray}
\epsilon _\Delta \simeq (4\sqrt{2}G_F)^{-1}\frac{Im(g_{\tau l}g_{l\tau
}^{*}) }{M_\Delta ^2} 
\end{eqnarray}
in triplet model respectively.

For the singlet model we assume that $f_{e\mu }$ is considerably smaller
than $f_{\tau l}$, so that one does not need to readjust the Fermi constant $%
G_F$. This assumption is also consistent with the constraint set by
universality between $\beta $ and $\mu $ decay \cite{Zee,Bri}. The parameter 
$\frac{Im(f_{\tau l}f_{l\tau }^{*})}{M_h^2}$ is constrained only by the
measurement of $\tau $ leptonic decays. At $2\sigma $ level (which is about $%
2\sim 3$\% precision) we estimate approximately $\frac{Im(f_{\tau l}f_{l\tau
}^{*})}{M_h^2}\leq 10^{-6}$ GeV$^{-2}$ \cite{Tao}. It implies that 
\begin{eqnarray}
\epsilon _h\simeq (4\sqrt{2}G_F)^{-1}\frac{Im(f_{\tau l}f_{l\tau }^{*})}{%
M_h^2}\leq 1.4\times 10^{-2} 
\end{eqnarray}
with $M_h=100$ GeV. Therefore in this model there is a possibility that CP/T
violation effect may show up with a size reachable at TCF in pure leptonic
decay channels.

For the triplet model the direct constraint is also from the measurement of
pure leptonic decays. The same result is obtained as that in the singlet
model , i.e. $\frac{Im(g_{\tau l}g_{l\tau }^{*})}{M_h^2}\leq 10^{-6}$ GeV$%
^{-2}$. As the result of this constraint one has 
\begin{eqnarray}
\epsilon _h\simeq (4\sqrt{2}G_F)^{-1}\frac{Im(g_{\tau l}g_{l\tau }^{*})}{%
M_\Delta ^2}\leq 1.4\times 10^{-2} 
\end{eqnarray}
with $M_\Delta =100$ GeV. However, in this model the new interactions will
induce lepton family number violating decay $\tau \to 3l$ and $\mu \to 3e$
through exchange of the double charged scalar particle $\Delta ^{--}$.
Without seeing any signal, one obtains some approximate bounds on the
coupling constants as the following \cite{Dat1} 
\begin{eqnarray}
\frac{|g_{\mu e}g_{ee}^{*}|}{M_\Delta ^2}\leq 5\times 10^{-12} 
\end{eqnarray}
and 
\begin{eqnarray}
\frac{|g_{\tau l}g_{ll}^{*}|}{M_\Delta ^2}\leq 10^{-8} 
\end{eqnarray}
for $M_\Delta =100$ GeV. If one assumes that all the couplings $g_{ij}$ are
at the same order of magnitude, then these bounds will restrict the CP/T
violation size far below the ability of TCF. Again we see some hierarchies
on the couplings are needed for this model to give rise observable CP/T
violation effects. Additionally in the triplet model one has to avoid the
restriction from neutrino mass generation \cite{Gel}. If neutrino develops a
mass at tree level, either the couplings or the vacuum expectation value of
the neutral component of the triplet $\Delta ^0$ are extremely small. The
natural way to deal with this problem is to impose some symmetry on this
model. An example is to introduce a discrete symmetry: 
\begin{eqnarray}
l\to il;\hskip 2cmE\to iE\hskip 2cm\Delta \to -\Delta 
\end{eqnarray}
With this symmetry, $\Delta ^0$ will never develop a nonzero vacuum
expectation value, therefore the couplings are not constrained by the
neutrino mass generation.

\section{Discussion and conclusion}

In this work we systematically investigate the possibility of finding CP/T
violation in the $\tau $ sector with TCF. The origin of CP/T violation is
from the extensions of the SM. We discuss most of the popular beyond the SM
and present the models which may give rise large CP/T violation in $\tau $
sector through either EDM or decay of $\tau $ lepton. Before making our
conclusion, some interesting points should be further discussed or
emphasized. (1) Polarization of initial electron and/or positron is very
desired for our purpose. First with polarization the precision of
measurement of EDM will be increased by about two order of magnitude, as $%
10^{-19}$e-cm, which is used through this work. Without polarization, from
our above discussion one sees that we have no hope to expect a detectable
EDM of $\tau $ at TCF. Secondly in some decay channels without final state
interaction, like pure leptonic decays and two body decays $\pi \nu _\tau $
etc., polarization is needed to search for CP/T violation occurring at $\tau 
$ decay vertex. With unpolarized electron and positron beams the CP/T
violation could only be detected using channels with final state interaction
phase, like $2\pi \nu _\tau $ etc. (2) For the hadronic decay we only
consider inclusive processes. The advantage of inclusive process is that one
does not need not to consider the hadronization of quarks, which may bring
in large uncertainties in the estimation. And the event number in inclusive
process is larger than that in certain exclusive processes. However we
should mention that for certain exclusive decays the CP/T violation
parameter $\epsilon $ can be larger than that in inclusive decay. One
example is from the multi-Higgs double model. We estimate that $\epsilon
\leq 3\times 10^{-4}$ for the decay $\tau \to {\bar u}d\nu _\tau $. Here we
may also consider the exclusive decay $\tau \to 3\pi \nu _\tau $ contributed
by $a_1$ and $\pi ^{\prime }$ resonances. Compared to inclusive decay, the $%
\epsilon $ parameter is larger by a factor of (using current algebra
relation) 
\begin{eqnarray}
\frac{<o|{\bar u}_Ld_R|\pi ^{\prime }>}{<o|{\bar u}_L\gamma _0d_L|\pi
^{\prime }>}\simeq \frac{m_{\pi ^{\prime }}}{m_u+m_d}\simeq 100.
\end{eqnarray}
So $\epsilon \leq 3\times 10^{-2}$ is obtained. However on the other hand
the event number decreases by a factor of 
\begin{eqnarray}
\frac{f_\pi ^{\prime }}{f_\pi }\frac{Br(\tau \to \pi \nu _\tau )}{Br(\tau
\to hadron+\nu _\tau )}\simeq 10^{-2}
\end{eqnarray}
Here $f_{\pi ^{\prime }}=5\times 10^3$ GeV is used. Therefore statistical
error increases by about 10 times. In the other words the measurement
precision at TCF for this channel is about $10^{-2}$. As the result, at $%
2\sigma $ level $\epsilon \simeq 3\times 10^{-2}$ is observable. This
estimation agrees with the exact result of reference \cite{Cho}. (3)
Obviously the numerical result we obtained above is quite crude. More
accurate estimation is necessary in the future. For instance through this
paper we assume that EDM as large as $10^{-19}$e-cm and $\epsilon $ as large
as $10^{-3}$ can be observed. This of course is a rough estimation. To be
more precise, Monte Carlo simulation is needed, which will tell us more
confidently how large CP/T violation is able to be observed at TCF.
Especially the Monte Carlo simulation on EDM of $\tau $ will give us a quite
clear result , because in this case the $d_\tau $ is the only parameter we
should take care. All the model dependence is included in it. Recently a 
group of people analyzed the data from BEPC experiments to 
set bound on the T-violating effect for $\tau$ system \cite{Qi}.  
Following the suggestion 
by T.D. Lee, they considered the pure leptonic $\tau^{\pm}$ decays to 
$e^{\pm}\mu^{\mp}$ plus neutrinos in the final states. The T-violating 
amplitude 
\begin{eqnarray}
A=<\hat{p}_e\cdot(\hat{p}_1\times \hat{p}_2)>_{average}
\end{eqnarray}
 is measured, where $\hat{p}_e$ is the unit momentum vector of the initial 
electron beam, $\hat{p}_1$ and $\hat{p}_2$ are the unit momenta of the final 
state electron and muon respectively.  Totally 251 events are analyzed and it 
results in 
\begin{eqnarray}
A=-0.097\pm 0.039\pm 0.135
\end{eqnarray}
This result agrees with no T-violation as expected from our previous 
discussion on pure leptonic $\tau$ decays.  
(4) In order to generate
detectable large CP/T violation effects, we know from our investigation that
there must exist new physics and the new physics scale is not far above the
weak scale. Therefore if there is a observable CP/T violation effect in $%
\tau $ sector at TCF, the associated new physics phenomena should be
observed at high energy experiments, like LHC and LEP II experiments. It is
interesting to see if the new particles predicted by the various models we
have discussed in this paper are indeed detectable in these high energy
experiments. (5) Precise measurement of the pure leptonic decay is another
way to test the new physics responsible for CP/T violation. Since if there
is CP/T violation effect at level of $10^{-3}$, the $\tau $ leptonic decay
width must deviate from the SM prediction at the same level. So we expect to
observe the deviation by measuring the branching ratio of the pure leptonic
decay. However it is not true {\it vice versa}, since a deviation of
leptonic branching ratio from that of the SM does not necessarily indicate
CP/T violation.

Finally we come to our conclusion. There exists the possibility that CP/T
violation in $\tau $ sector is large enough to be discovered at TCF,
although for this large violation effect some specific new physics phenomena
beyond the SM are needed and the parameter spaces of the models are strongly
restricted.

Z. J. Tao is supported by the National Science Foundation of China (NSFC).

\newpage 
\centerline{Figure Caption} Fig.1 One loop diagram for lepton EDM
generation, where $F$ is heavy fermion and $S$ is the new boson. Photon 
line is attached to charged particles in the loop. 

Fig.2 The diagrams for $\tau$ decay. (a) is the contribution from the SM and
(b) is the contribution from new boson exchange.

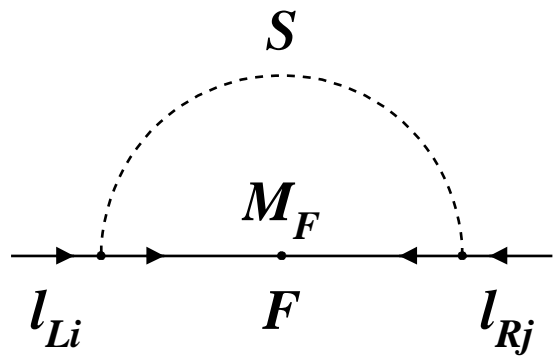
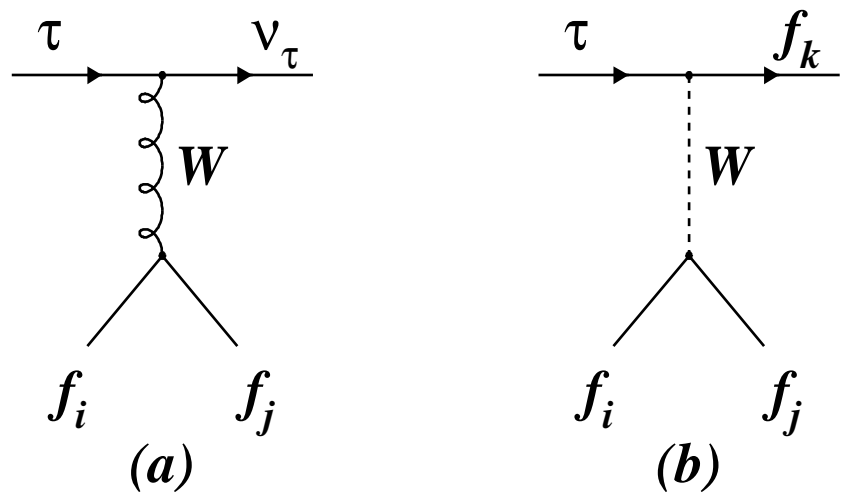
\end{document}